\documentclass{revtex4}

\usepackage[usenames,dvipsnames]{color}
\usepackage{amsfonts,amssymb,amsmath,amsthm}
\usepackage{bm}
\usepackage[a4paper]{geometry}
\usepackage{pdflscape}
\usepackage{graphicx}
\usepackage{mathrsfs}
\usepackage{hyperref}

\definecolor{darkred}{rgb}{.8,0,0}

\definecolor{darkblue}{rgb}{0,0,.7}

\newcommand{\eps}{\varepsilon}

\newcommand{\gr}[2]{\langle #1 \rangle_{#2}}
\newcommand{\scal}[1]{\langle #1 \rangle}
\newcommand{\rev}[1]{\widetilde{#1}}
\newcommand{\blade}[2]{{#1}_1 \wedge \ldots \wedge {#1}_{#2}}
\newcommand{\revblade}[2]{{#1}_{#2} \wedge \ldots \wedge {#1}_{1}}

\theoremstyle{plain}
\newtheorem*{VarPrinc}{Variational principle}
\newtheorem*{CanEOM}{Canonical equations of motion}

\begin{document}

\title{Classical field theories from Hamiltonian constraint: \\
Canonical equations of motion and local Hamilton-Jacobi theory}

\author{V\'{a}clav Zatloukal}

\email{zatlovac@fjfi.cvut.cz}

\homepage{http://www.zatlovac.eu}

\affiliation{\vspace{3mm}
Faculty of Nuclear Sciences and Physical Engineering, Czech Technical University in Prague, \\
B\v{r}ehov\'{a} 7, 115 19 Praha 1, Czech Republic \\
}

\affiliation{
Max Planck Institute for the History of Science, Boltzmannstrasse 22, 14195 Berlin, Germany
}

\begin{abstract}
Classical field theory is considered as a theory of unparametrized surfaces embedded in a configuration space, which accommodates, in a symmetric way, spacetime positions and field values. Dynamics is defined by a (Hamiltonian) constraint between multivector-valued generalized momenta, and points in the configuration space. Starting from a variational principle, we derive local equations of motion, that is, differential equations that determine classical surfaces and momenta. A local Hamilton-Jacobi equation applicable in the field theory then follows readily. The general method is illustrated with three examples: non-relativistic Hamiltonian mechanics, De Donder-Weyl scalar field theory, and string theory.

\end{abstract}

\maketitle

\section{Introduction}

In non-relativistic mechanics, the trajectory of a particle is a function $x(t)$, which describes how the position of the particle changes with time.
In relativistic mechanics, space and time are treated equally, and the particle's trajectory is regarded as a sequence of spacetime points $(t,x)$. 

In field theory, the field configuration is usually viewed as a function $\phi(x)$ that describes how the field varies from point to point. However, general relativity suggests \cite{RovelliQG} that the spacetime is a dynamical entity, which should be put with fields on the same footing. Mathematically, instead of a function $\phi(x)$ one should therefore consider the respective graph, i.e., the collection of points $(x,\phi)$.

In this article, we study the mathematical formalism proposed in \cite[Ch.~3]{RovelliQG} that treats time, space, and fields equally. All these entities are collectively called \emph{partial observables}, and together they form a finite-dimensional \emph{configuration space}. Classical field theory predicts that certain correlations between partial observables can be realized in nature. These are then called \emph{physical motions}, and have the form of surfaces embedded in the configuration space.

Our dynamical description utilizes multivector-valued momentum variable, which can be thought of as conjugated to the motion's tangent planes; thus generalizing the canonical momentum conjugated to the velocity vector in classical mechanics. Individual theory is specified by a choice of the \emph{Hamiltonian} $H$, which is a function of a configuration space point $q$ and momentum $P$. This Hamiltonian enters into a variational principle (Section \ref{sec:VarPrinc}) via the \emph{Hamiltonian constraint} $H(q,P) = 0$.

The aim of this article is to establish, in the first place, equations of motion that follow from the variational principle. This is done in Section \ref{sec:CanEq}, Eqs. (\ref{CanEOM}). These equations generalize the Hamilton's canonical equations of motion of classical mechanics. From Eqs. (\ref{CanEOM}) we derive the local Hamilton-Jacobi equation (\ref{HJeq}), which generalizes to the field theory the respective concept from classical mechanics (in this context, see also Refs. \cite{Kastrup} and \cite{Rund}). It should be stressed that both, Eqs. (\ref{CanEOM}) and (\ref{HJeq}), contain only partial, not variational, derivatives. 

Three examples are provided in Section \ref{sec:Examples} to demonstrate universality of the present formalism. The first example (\ref{sec:ExNonRel}) shows how non-relativistic mechanics is deduced when we assume that the motions are one dimensional curves, and choose the Hamiltonian $H$ appropriately. Eqs. (\ref{CanEOM}) then reduce to the Hamilton's canonical equations, accompanied by the law of energy conservation. The Hamilton-Jacobi equation of classical mechanics is also recovered. 

The second example (\ref{sec:ExScalar}) discusses the theory of real one-component scalar field defined on a Euclidean spacetime of any dimension. It is shown that Eqs. (\ref{CanEOM}) produce the De Donder-Weyl equations \cite{DeDonder,Weyl,Kanat1999}, and, at the same time, they incorporate the continuity equation for the energy-momentum tensor. Hamilton-Jacobi equation reproduces the one invented by Weyl \cite{Weyl}.

In the last example (\ref{sec:ExString}) we treat relativistic particle, string, or higher-dimensional membrane, depending on the dimensionality of the motions. The configuration space is identified with the target space of the string theory, motions are the worldsheets, and the corresponding Hamiltonian is essentially the simplest and most symmetric function of the momentum variable. The equations of motion have simple geometric meaning, namely, they ensure that the mean curvature of the physical motion vanishes. In fact, this is exactly the condition that defines \emph{minimal surfaces} \cite{Osserman}. 

One more remark is in order before we start. All manipulations are performed in the mathematical formalism of geometric (or Clifford) algebra and calculus developed by D. Hestenes \cite{Hestenes}. It is a coordinate-free language that is more universal than the calculus of differential forms, nevertheless, it is yet not well-recognized by a broad audience. Reader unfamiliar with geometric algebra or calculus is recommended to first read Appendix \ref{sec:GAGC}, where we introduce the basics, and derive or quote some key results that are used in the main text.

\section{Variational principle} \label{sec:VarPrinc}

We start with a set of partial observables that constitute a $(D+N)$-dimensional Euclidean configuration space $\mathcal{C}$. A point $q$ in the configuration space represents simultaneous measurement of all partial observables, e.g., $q=(x,\phi)$.
To establish a \emph{physical} theory, one has to specify correspondence between the partial observables, and physical measuring devices, such as clock, rulers, or instruments measuring components of the field. In this article we take such correspondence for granted, as we are only concerned with the \emph{mathematical} aspects of the theory.

Denote by $D$ the dimensionality of motions, i.e., submanifolds $\gamma$ of the configurations space $\mathcal{C}$. With $D=1$ we can study particle mechanics, with $D=2$ we can do string theory, or field theory in two dimensions, and so on. We shall not deal with systems with gauge invariance, for which the mathematical motion (the surface in $\mathcal{C}$) has higher dimensionality than the actual physical motion (the trajectory).

Tangent space of $\gamma$ at point $q$ is spanned by $D$ linearly independent vectors $a_1,\ldots,a_D$, which are conveniently combined into a grade-$D$ multivector $\blade{a}{D}$. Normalized version of the latter is called the \emph{unit pseudoscalar of $\gamma$}, and it is denoted by $I_\gamma$. In the terminology used in Ref. \cite[Ch.~6]{Frankel}, the function $I_\gamma(q)$ represents a $D$-dimensional distribution on $\mathcal{C}$, with $\gamma$ being its integral manifold.

\begin{figure} 
\includegraphics[scale=1]{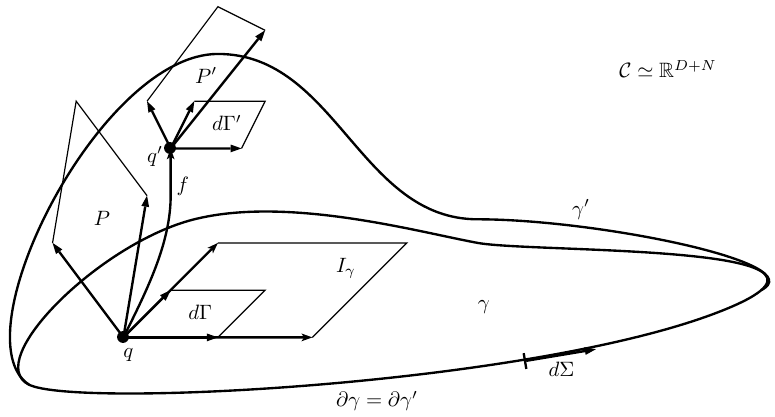}
\caption{Variational principle.}
\label{fig:VarPrinc}
\end{figure}

Fundamental for the following formulation of dynamics is the concept of generalized momentum, which is a grade-$D$ multivector, denoted by $P$, defined at each point of $\gamma$ (see Fig.~\ref{fig:VarPrinc}). It serves as a quantity conjugated to $I_\gamma$, thus generalizing the canonical momentum of particle mechanics.

The last ingredient is the Hamiltonian $H(q,P)$, which is supposed to be scalar-valued. (Generalization to the case of multicomponent $H$ is straightforward.) 

The variational principle can now be stated as follows (cf. \cite[Ch.~3.3.2]{RovelliQG}): 
\begin{VarPrinc}
A surface $\gamma_{\rm cl}$ with boundary $\partial \gamma_{\rm cl}$ is a physical motion, if the couple $(\gamma_{\rm cl},P_{\rm cl})$ extremizes the (action) functional
\begin{equation} \label{Action}
\mathcal{A}[\gamma,P] = \int_\gamma P(q) \cdot d\Gamma(q)
\end{equation}
in the class of pairs $(\gamma,P)$, for which $\partial\gamma = \partial\gamma_{\rm cl}$, and $P$ defined along $\gamma$ satisfies
\begin{equation} \label{HamConstraint}
H(q,P(q)) = 0 ~~~~~ \forall q \in \gamma .
\end{equation}
\end{VarPrinc}

(The subscript ``cl" stands for ``classical" motion, or trajectory, which we sometimes use instead of the expression ``physical" motion.)

The integral in (\ref{Action}) is defined in (\ref{GCintGen}) without having recourse to a parametrization of the surface $\gamma$. 
Of course, if desired, the oriented surface element $d\Gamma$ can be cast, using arbitrary coordinates on $\gamma$, as
\begin{equation} \label{SurfElCoords}
d\Gamma 
= \left( \frac{\partial q}{\partial \tau_1} \wedge \ldots \wedge \frac{\partial q}{\partial \tau_D} \right) d\tau_1 \ldots d\tau_D .
\end{equation}

The integrand in (\ref{Action}) may be rephrased as a differential form $\theta = p_{j_1 \ldots j_D} dq^{j_1} \wedge \ldots \wedge dq^{j_D}$ (see Formula (\ref{GAdifForm}) in Appendix \ref{sec:GAGC}). However, the merit of geometric calculus consists, in fact, in splitting of the differential form into two parts, $d\Gamma$ and $P$, where $P$ is able to enter into functions such as $H(q,P)$. 

Finally, let us note that in \cite[Ch.~3.3.2]{RovelliQG} the action is an integral of $\theta$ over the submanifolds of the bundle of $D$-forms over $\mathcal{C}$. Since we hesitate to work in spaces mixing points $q$ and multivectors $P$, we rather operate with surfaces in $\mathcal{C}$, on which the momentum field is defined.


\section{Canonical equations of motion} \label{sec:CanEq} 

We will now derive the equations of motion that follow from the variational principle of Section~\ref{sec:VarPrinc}. For this purpose, we incorporate the Hamiltonian constraint (\ref{HamConstraint}) into the action (\ref{Action}) by means of a scalar Lagrange multiplier $\lambda$. The augmented action is the functional
\begin{equation} \label{ActionAugm}
\mathcal{A}[\gamma,P,\lambda] 
= \int_\gamma \left[ P(q) \cdot d\Gamma(q) - \lambda(q) H(q,P(q)) \right] ,
\end{equation}
where $\lambda$ is in fact an infinitesimal quantity with magnitude comparable to $|d\Gamma|$ -- the magnitude of $d\Gamma$ (see definition (\ref{GAmagnitude})).

Varied action $\mathcal{A}[\gamma',P',\lambda']$ is the integral taken over a new surface $\gamma'$, and featuring new functions $P'$ and $\lambda'$, which are defined along $\gamma'$ (see Fig.~\ref{fig:VarPrinc}). Let $f$, where 
\begin{equation}
f(q) = q + \delta q(q) ,
\end{equation}
be the infinitesimal diffeomorphism mapping between $\gamma$ and $\gamma'$, i.e., $\gamma'=\{q'=f(q)\,|\,q \in \gamma \}$, and denote by 
\begin{equation}
\delta P(q) \equiv P'(f(q)) - P(q) 
~~~~~{\rm and}~~~~~ 
\delta \lambda(q) \equiv \lambda'(f(q)) - \lambda(q)
\end{equation}
the variations of momentum and Lagrange multiplier, respectively.

Variation of the action (\ref{ActionAugm}), $\delta\mathcal{A} \equiv \mathcal{A}[\gamma',P',\lambda'] - \mathcal{A}[\gamma,P,\lambda]$, is given by
\begin{equation}
\delta\mathcal{A} 
= \int_\gamma \left[ P'(f(q)) \cdot \underline{f}(d\Gamma(q);q) \!-\! \lambda'(f(q)) H \big(f(q),P'(f(q)) \big) \right]
- \int_\gamma \left[ P(q) \cdot d\Gamma(q) \!-\! \lambda(q) H(q,P(q)) \right] ,
\end{equation}
where we have employed the integral substitution theorem (\ref{GCintChangeVar}) to transform the integral over $\gamma'$ into an integral over $\gamma$. For the infinitesimal diffeomorphism $f$, the outermorphism mapping $\underline{f}$ that specifies the transformation rule for multivectors, is given by Formula (\ref{GAinfsmOut}). Therefore, up to first order in $\delta q$, $\delta P$ and $\delta \lambda$, we find
\begin{align} \label{ActionVar}
\delta\mathcal{A}
&= \int_\gamma \left[ 
(P + \delta P) \cdot \big(d\Gamma + (d\Gamma \cdot \partial_q) \wedge \delta q \big) - (\lambda + \delta\lambda) H(q + \delta q, P + \delta P)
- P \cdot d\Gamma + \lambda\, H(q,P) 
\right]
\nonumber\\
&\approx 
\int_\gamma \left[
- \delta\lambda \, H(q,P) 
+ \delta P \cdot \big(d\Gamma - \lambda \, \partial_P H(q,P) \big)
- \lambda \, \delta q \cdot \dot{\partial}_q H(\dot{q},P)
+ P \cdot \big((d\Gamma \cdot \partial_q) \wedge \delta q \big)
\right] ,
\end{align}
where the \emph{vector derivative} $\partial_q$, and the \emph{multivector derivative} $\partial_P$ are defined in (\ref{GAvectDer}), and (\ref{GCmultivecDeriv}), respectively.
The ``overdot" notation is used here to indicate the scope of the differential operator $\partial_q$, and has nothing to do with time derivative. Without an overdot, any differential operator is supposed to act on functions that stand to its right.

The last term in (\ref{ActionVar}) can be recast with a help of the \emph{Fundamental theorem of geometric calculus} (\ref{GCintFundThm}),
\begin{equation} \label{PerPartes}
\int_\gamma P \cdot \big((d\Gamma \cdot \partial_q) \wedge \delta q \big)
= \int_{\partial\gamma} P \cdot (d\Sigma \wedge \delta q)
- \int_\gamma \dot{P} \cdot \big((d\Gamma \cdot \dot{\partial}_q) \wedge \delta q \big) ,
\end{equation}
where $d\Sigma$ is the oriented volume element of the boundary $\partial\gamma$. Now, the first term on the right-hand side vanishes, since we assume that $\gamma$ and $\gamma'$ have common boundary. 

For $D=1$, $d\Gamma \cdot \partial_q$ is algebraically a scalar, and so the integrand in the second term is readily reshuffled,
\begin{equation}
\dot{P} \cdot \big((d\Gamma \cdot \dot{\partial}_q) \wedge \delta q \big)
= \delta q \cdot \, (d\Gamma \cdot \partial_q P) .
\end{equation}
(Mind the priority of the inner product ``$\cdot$", and the outer product ``$\wedge$" before the geometric product, which is denoted by an empty symbol.)

For $D>1$, we can use identities (\ref{GAidentity4}) and (\ref{GAidentity1}) to find
\begin{equation}
\dot{P} \cdot \big((d\Gamma \cdot \dot{\partial}_q) \wedge \delta q \big)
= \big( \dot{P} \cdot (d\Gamma \cdot \dot{\partial}_q) \big) \cdot \delta q
= (-1)^{D-1} \delta q \cdot \big( (d\Gamma \cdot \partial_q) \cdot P \big) .
\end{equation}
The two cases have to be treated separately due to the definition (\ref{GAinnerouter}) of the inner product.

After these rearrangements we arrive at our final expression for the variation of the action,
\begin{equation}
\delta\mathcal{A}
\approx 
\int_\gamma \left[
- \delta\lambda \, H(q,P) 
+ \delta P \cdot \big( d\Gamma - \lambda \, \partial_P H(q,P) \big)
+ \delta q \cdot \left(
(-1)^D (d\Gamma \cdot \partial_q) \cdot P
- \lambda \, \dot{\partial}_q H(\dot{q},P)
 \right)
\right] ,
\end{equation}
which holds for $D>1$, while the case $D=1$ is obtained simply by replacing $(d\Gamma \cdot \partial_q) \cdot P$ with $d\Gamma \cdot \partial_q P$.
The requirement that the variation of the action be zero for all $\delta P$, $\delta q$ and $\delta\lambda$ yields the following
\begin{CanEOM}
Physical motions $\gamma_{\rm cl}$ are obtained by solving the system of equations
\begin{subequations} \label{CanEOM}
\begin{align} 
\label{CanEOM1}
\lambda \, \partial_P H(q,P) &= d\Gamma , 
\\ \label{CanEOM2}
(-1)^D \lambda \, \dot{\partial}_q H(\dot{q},P) 
&= \begin{cases}
d\Gamma \cdot \partial_q P & ~~{\rm for}~ D=1 \\
(d\Gamma \cdot \partial_q) \cdot P &  ~~{\rm for}~ D>1 ,
\end{cases}
\\
\label{CanEOM3} 
H(q,P) &= 0 .
\end{align}
\end{subequations}
\end{CanEOM}
(We use the adjective ``canonical", because these equations generalize, as we shall see in Example~\ref{sec:ExNonRel}, Hamilton's canonical equations of motion of classical mechanics.)

The first canonical equation (\ref{CanEOM1}) furnishes a relation between the momentum $P$, and the tangent planes of $\gamma$ represented by the oriented surface element $d\Gamma$. It asserts that the multivector derivative $\partial_P H$, which is a grade-$D$ multivector, is proportional to $d\Gamma$, with the proportionality constant equal to $\lambda$. Note that one can always normalize $d\Gamma$ and $\lambda$ by the magnitude $|d\Gamma|$ to free Eqs.~(\ref{CanEOM}) from infinitesimal quantities.

The second canonical equation (\ref{CanEOM2}) describes how the momentum multivector $P$ changes as it slides along the surface $\gamma$. It is important to note that $P$ is differentiated effectively only in the directions parallel to $\gamma$, as a consequence of the inner product between the surface element $d\Gamma$, and the vector derivative $\partial_q$. Moreover, the ``overdot" on the left-hand side  assures that only explicit dependence of $H$ on $q$ is being differentiated, not the dependence through $P(q)$.

The last canonical equation (\ref{CanEOM3}) is the Hamiltonian constraint (\ref{HamConstraint}). Let us remark that had we started with several constraints $H_j(q,P)=0$ in the variational principle, we would have introduced the corresponding number of Lagrange multipliers $\lambda_j$, and the canonical equations would contain the terms $\sum_j\lambda_jH_j$ instead of $\lambda\,H$.

In Appendix~\ref{sec:Comp} we provide a component form of the canonical equations to make them more accessible for a reader who is not sufficiently familiar with the formalism of geometric algebra. Note, however, that this step is not necessary in order to make practical calculations, as will be illustrated in the examples of Sec.~\ref{sec:Examples}.

\section{Local Hamilton-Jacobi theory} \label{sec:LocHJ}

One possible method to approach the canonical equations (\ref{CanEOM}) is the following. Suppose $P(q)$ is given, that obeys the Hamiltonian constraint
\begin{equation}
H(q,P(q)) = 0 
\end{equation}
on some open subset of the configuration space $\mathcal{C}$. By differentiation, we obtain, according to the chain rule (\ref{GCchainH}),
\begin{equation}
\dot{\partial}_q H(\dot{q},P(q))
+ \dot{\partial}_q \dot{P}(q) \cdot \partial_P H(q,P(q)) = 0 ,
\end{equation}
and using the first canonical equation (\ref{CanEOM1}), we find
\begin{equation}
\lambda \, \dot{\partial}_q H(\dot{q},P(q))
= - \dot{\partial}_q \dot{P}(q) \cdot d\Gamma .
\end{equation}
But the right-hand side may be rearranged by means of identity (\ref{GAexpansion}) for $D=1$, or (\ref{GAidentity1}) and (\ref{GAidentity7}) for $D>1$, with the result
\begin{equation}
\lambda \, \dot{\partial}_q H(\dot{q},P(q)) =
\begin{cases}
d\Gamma \cdot \big( \partial_q \wedge P(q) \big)
- d\Gamma \cdot \partial_q P(q) & ~~{\rm for}~ D=1
\\
(-1)^{D-1} d\Gamma \cdot \big( \partial_q \wedge P(q) \big)
+ (-1)^D (d\Gamma \cdot \partial_q) \cdot P(q) & ~~{\rm for}~ D>1 .
\end{cases}
\end{equation}

Therefore, we observe that if
\begin{equation}
\partial_q \wedge P(q) = 0 ,
\end{equation}
the second canonical equation (\ref{CanEOM2}) is automatically fulfilled. Momentum satisfying this condition can be expressed, at least locally, as $P(q) = \partial_q \wedge S(q)$, where $S$ is a multivector of grade $D-1$ (consider relation (\ref{GCformExtDeriv})). Canonical equations (\ref{CanEOM}) are then reduced to two equations,
\begin{equation} \label{HJeqGamma}
\lambda \, \partial_P H(q,\partial_q \wedge S) = d\Gamma ,
\end{equation}
and the \emph{local Hamilton-Jacobi equation}
\begin{equation} \label{HJeq}
H(q,\partial_q \wedge S) = 0 
\end{equation}
(see Appendix~\ref{sec:Comp} for a component form of this equation).
If we succeed in finding a solution of Eq. (\ref{HJeq}), we can plug it into Eq. (\ref{HJeqGamma}), which then defines a distribution of tangent planes of a classical motion surface. This distribution is integrable only if certain conditions are satisfied (see \cite[Ch.~6.1]{Frankel}).

In addition, if we find a whole family of solution $S(q;\alpha)$, parametrized by a continuous parameter $\alpha$, then differentiating Eq. (\ref{HJeq}) with respect to $\alpha$, and substituting Eq. (\ref{HJeqGamma}), yields
\begin{equation} \label{HJparamSol}
0 = \lambda \, \partial_\alpha H(q,\partial_q \wedge S) 
= \lambda \, \dot{\partial}_\alpha (\partial_q \wedge \dot{S}) \cdot \partial_P H(q,\partial_q \wedge S)
= d\Gamma \cdot \big(\partial_q \wedge (\partial_\alpha S) \big) .
\end{equation}
Now, for $D=1$, the Hamilton-Jacobi function $S$ is scalar-valued, and we have
\begin{equation} \label{HJconserved}
d\Gamma \cdot \partial_q (\partial_\alpha S) = 0 
~~~\Rightarrow~~~
\partial_\alpha S(q;\alpha) = \beta ~~~~~\forall q \in \gamma_{\rm cl} ,
\end{equation}
for some constant $\beta$, meaning that the quantity $\partial_\alpha S(q;\alpha)$ is conserved along physical motion. If one finds $N$ such parameters (recall that the dimension of the configuration space is $N+1$), the physical motion $\gamma_{\rm cl}$ can be given implicitly by a set of constraints between partial observables,
\begin{align}
\partial_{\alpha_1} S(q;\alpha_1,&\ldots,\alpha_N) = \beta_1 \nonumber \\
&\vdots \nonumber \\
\partial_{\alpha_N} S(q;\alpha_1,&\ldots,\alpha_N) = \beta_N .
\end{align}
Of course, we assume that the $N$ constraints are independent, i.e., that the gradients $\partial_q (\partial_{\alpha_1} S), \ldots, \partial_q (\partial_{\alpha_N} S)$ are at each point linearly independent. In Example \ref{sec:ExString} we will illustrate the Hamilton-Jacobi method with the case of a relativistic particle. 

When $D>1$, we can use identity (\ref{GAidentity4}) and theorem (\ref{GCintFundThm}) to cast Eq. (\ref{HJparamSol}) as
\begin{equation} \label{HJcontEq}
(d\Gamma \cdot \partial_q) \cdot (\partial_\alpha S) = 0 
~~~\Rightarrow~~~
\int_{\bar{\gamma}_{\rm cl}} (d\Gamma \cdot \partial_q) \cdot (\partial_\alpha S)
= \int_{\partial \bar{\gamma}_{\rm cl}} d\Sigma \cdot (\partial_\alpha S)
= 0 ,
\end{equation}
where $\bar{\gamma}_{\rm cl}$ is an arbitrary $D$-dimensional subset of $\gamma_{\rm cl}$ (a ``patch" on $\gamma_{\rm cl}$). Therefore, in the multidimensional case, the conservation law (\ref{HJconserved}) is replaced with a certain continuity equation.

One remark is in order before closing this section. In classical particle mechanics, one of the solutions of the Hamilton-Jacobi equation is the action along classical trajectory, regarded as a function of one of the endpoints. In field theory, the classical action may be viewed as a functional of the boundary $\partial\gamma_{\rm cl}$. Some authors (e.g. \cite[Ch.~3.3.4]{RovelliQG}) have therefore considered a variational differential equation that describes how the classical action changes under variations of the boundary, naming it also ``Hamilton-Jacobi equation". Note that Eq. (\ref{HJeq}) is substantially different from this kind of approach, for it contains only partial, not variational, derivatives. That is why we call it ``local Hamilton-Jacobi equation". Local Hamilton-Jacobi theory is also treated, e.g., in Refs. \cite{Kastrup} and \cite{Rund}.

\section{Examples} \label{sec:Examples}

In the following examples, we illustrate the general theory by specifying a concrete form of the Hamiltonian $H(q,P)$. Reader's familiarity with the techniques of geometric algebra and calculus on the level of Appendix \ref{sec:GAGC} is assumed.

\subsection{Non-relativistic Hamiltonian mechanics} \label{sec:ExNonRel}

\begin{figure} 
\includegraphics[scale=1]{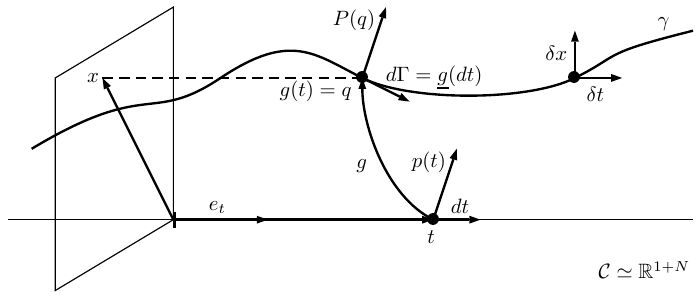}
\caption{Non-relativistic Hamiltonian mechanics.}
\label{fig:NonRel}
\end{figure}

Let us consider $D=1$, and choose a constant unit vector $e_t$ in the configuration space $\mathcal{C}~\simeq~\mathbb{R}^{1+N}$. Arbitrary point $q$ can be decomposed as $q = t+x$, where $t$ is parallel to $e_t$, while $x$ is perpendicular to $e_t$ (see Fig.~\ref{fig:NonRel}). 

Define the Hamiltonian as follows:
\begin{equation} \label{NonRelHamConstr}
H(q,P) = P \cdot e_t + H_0(q,P) ,
\end{equation}
where $e_t \cdot \partial_P H_0 = 0$. We shall identify $H_0$ as the non-relativistic Hamiltonian of a mechanical system.

``Dotting" the first canonical equation (\ref{CanEOM1}) with $e_t$, we find
\begin{equation} \label{NonRelLambda}
e_t \cdot d\Gamma = \lambda\, e_t \cdot \partial_P H = \lambda \neq 0 .
\end{equation}
This means that the tangent vector of a physical motion $\gamma_{\rm cl}$ has nonvanishing component parallel to $e_t$, and we can therefore present $\gamma_{\rm cl}$ as
\begin{equation}
\gamma_{\rm cl} = 
\{ q = g(t) = t + x(t) \,|\, t \in {\rm span}\{ e_t \} \simeq \mathbb{R} \} .
\end{equation}
The line element on the $t$-axis is related to the line element on $\gamma_{\rm cl}$ via the differential mapping (\ref{GAdifMap}),
\begin{equation}
d\Gamma = \underline{g}(dt) = dt + dt \cdot \partial_t x .
\end{equation}
Eq.~(\ref{NonRelLambda}) then implies
\begin{equation}
\lambda = e_t \cdot d\Gamma = e_t \cdot dt = e_t\,dt 
~~~\Rightarrow~~~ dt = \lambda\, e_t 
~~~\Rightarrow~~~ d\Gamma = \lambda\, \underline{g}(e_t) = \lambda (e_t + e_t \cdot \partial_t x) ,
\end{equation}
which allows us to eliminate $\lambda$ from the equations of motion.

Let us denote $p(t) \equiv P(g(t))$, and observe that the first canonical equation (\ref{CanEOM1}) assumes the form
\begin{equation} \label{NonRelHamEq1}
e_t \cdot \partial_t x = \partial_p H_0(q,p) .
\end{equation}

As concerns the second canonical equation (\ref{CanEOM2}), we realize that by the chain rule for differentiation
\begin{equation}
e_t \cdot \partial_t p(t) 
= e_t \cdot \partial_t P(g(t)) 
= \underline{g}(e_t) \cdot \partial_q P(q)|_{q=g(t)}
= \frac{1}{\lambda} d\Gamma \cdot \partial_q P(q)|_{q=g(t)} ,
\end{equation}
and so we arrive at
\begin{equation} \label{NonRelEq2}
e_t \cdot \partial_t p = - \partial_q H(q,p) = - \partial_q H_0(q,p) .
\end{equation}

Projecting Equation (\ref{NonRelEq2}) onto $e_x$, an arbitrary vector perpendicular to $e_t$, we find
\begin{equation} \label{NonRelHamEq2}
e_t \cdot \partial_t \, e_x \cdot p 
= - e_x \cdot \partial_q H_0(q,p) ,
\end{equation}
while projecting onto $e_t$ yields
\begin{equation} \label{NonRelHamEq2t}
e_t \cdot \partial_t \, e_t \cdot p = 
- e_t \cdot \partial_q H_0(q,p) .
\end{equation}
Using finally the Hamiltonian constraint, Eq.~(\ref{CanEOM3}), with $H$ given by (\ref{NonRelHamConstr}), Eq. (\ref{NonRelHamEq2t}) implies
\begin{equation} \label{NonRelEnergyCons}
e_t \cdot \partial_t H_0(q(t),p(t))
= e_t \cdot \partial_q H_0(q,p(t))|_{q=g(t)} .
\end{equation}

It is now easy to realize that Eqs. (\ref{NonRelHamEq1}) and (\ref{NonRelHamEq2}) represent Hamilton's canonical equations of motion for a non-relativistic system with non-relativistic Hamiltonian $H_0$, while Eq. (\ref{NonRelEnergyCons}) expresses the law of conservation of the total energy $H_0$, if $e_t \cdot \partial_q H_0 = 0$, i.e., if $H_0(q,p)$ does not depend explicitly on time. Intuitively, the Hamilton's canonical equations follow from variations of the trajectory $\gamma$ in the $x$-space, while the energy conservation is a result of variations in the $t$-space (see Fig. \ref{fig:NonRel}).

Hamilton-Jacobi equation (\ref{HJeq}) for a scalar function $S(q)$ reads
\begin{equation}
H(q,\partial_q S) = e_t \cdot \partial_q S + H_0(q,\partial_q S) = 0 ,
\end{equation}
and hence reproduces the standard Hamilton-Jacobi equation of classical mechanics.

\subsection{Scalar field theory} \label{sec:ExScalar}

In this example we will show that the formalism based on Hamiltonian constraint can accommodate the theory of real one-component scalar field $\phi(x)$, defined on a $D$-dimensional Euclidean spacetime by the Lagrangian
\begin{equation} \label{ScFieldLagr}
\mathcal{L}(\phi,\partial_x \phi) = \frac{1}{2} (\partial_x \phi)^2 - V(\phi) .
\end{equation}

\begin{figure} 
\includegraphics[scale=1]{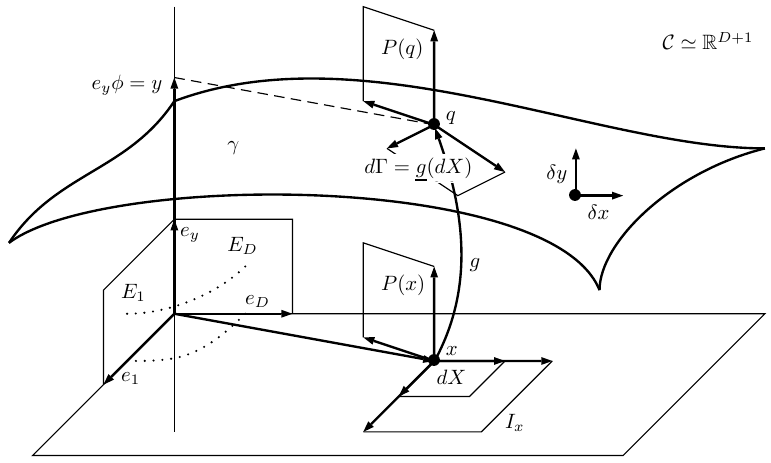}
\caption{Scalar field theory.}
\label{fig:ScField}
\end{figure}

For this purpose, let us assume $D>1$, choose a unit $D$-blade $I_x$ in a ($D+1$)-dimensional configuration space $\mathcal{C}$, and define the Hamiltonian
\begin{equation} \label{ScFieldHamConstr}
H(q,P) = P \cdot I_x + H_{DW}(q,P) ,
\end{equation}
where $I_x \cdot \partial_P H_{DW} = 0$. The blade $I_x$ defines a splitting of the configuration space $\mathcal{C}$ into a $D$-dimensional spacetime, spanned by an orthonormal set of vectors $\{e_1,\ldots,e_D\}$, $I_x = e_1 \ldots e_D$, and the field space, which is its one-dimensional orthogonal complement, represented by a unit vector $e_y$ (see Fig.~\ref{fig:ScField}).

In analogy with Example \ref{sec:ExNonRel}, let us take the inner product of the first canonical equation (\ref{CanEOM1}) with $\rev{I}_x$, the reversion of $I_x$:
\begin{equation} \label{ScFieldLambda}
\rev{I}_x \cdot d\Gamma = \lambda \rev{I}_x \cdot \partial_P H 
= \lambda \rev{I}_x \cdot I_x = \lambda \neq 0 .
\end{equation}
This implies for any vector $a$ tangent to $\gamma_{\rm cl}$ (i.e., $a \wedge d\Gamma = 0$) that
\begin{equation}
0 \neq \rev{I}_x \cdot d\Gamma 
=
\rev{I}_x \cdot [ a \wedge (a^{-1} \cdot d\Gamma) ] 
=
(\rev{I}_x \cdot a) \cdot (a^{-1} \cdot d\Gamma)
~~~\Rightarrow ~~~
I_x \cdot a \neq 0 .
\end{equation}
This means that no tangent vector is perpendicular to $I_x$, and we can therefore assume that a physical motion $\gamma_{\rm cl}$ is presented as 
\begin{equation}
\gamma_{\rm cl} = \{ q = g(x) = x+y(x) \,|\, x \in {\rm span}\{ e_1,\ldots,e_D \} \simeq \mathbb{R}^D \} .
\end{equation}

The surface element $d\Gamma$ is related to the infinitesimal element of the $x$-space $dX = |dX| I_x$ via the outermorphism mapping induced by the function $g$, $d\Gamma = \underline{g}(dX)$. It is not hard to show, in analogy with Equations (\ref{GAinfsmOutDeriv}) and (\ref{GAinfsmOut}), that for any $r$-vector $A_r$ from the $x$-space,
\begin{equation} \label{ScFieldDifMap}
\underline{g}(A_r) = A_r + (A_r \cdot \partial_x) \wedge y ,
\end{equation}
where $\partial_x = \sum_{j=1}^D e_j e_j \cdot \partial_q$.
Eq. (\ref{ScFieldLambda}) then supplies the relation
\begin{equation}
\lambda = \rev{I}_x \cdot d\Gamma = \rev{I}_x dX
~~~\Rightarrow~~~ dX = \lambda I_x
~~~\Rightarrow~~~ d\Gamma = \lambda \underline{g}(I_x) = \lambda \big(I_x + (I_x \cdot \partial_x) \wedge y \big) .
\end{equation}

The first canonical equation (\ref{CanEOM1}) reduces to
\begin{equation} \label{ScFieldEq1}
(I_x \cdot \partial_x) \wedge y = \partial_P H_{DW} .
\end{equation}

As concerns the second canonical equation, we observe that by the chain rule for differentiation,
\begin{equation}
e_j \cdot \partial_x P(g(x)) = \underline{g}(e_j) \cdot \partial_q P(q)|_{q = g(x)} = e_j \cdot \overline{g}(\partial_q) P(q)|_{q = g(x)}
~~~\Rightarrow~~~ \partial_x = \overline{g}(\partial_q) ,
\end{equation}
where $\overline{g}$ is the adjoint of the linear mapping $\underline{g}$ (see Eq. (\ref{GCadjoint})). With a usual abuse of notation, we will regard $P$ as a function of $x$, $P(x) \equiv P(g(x))$, from now on. Using the rules (\ref{GAadjointIdent}), we can cast the left-hand-side of the second canonical equation (\ref{CanEOM2}) as $\underline{g}(I_x \cdot \partial_x) \cdot P$, which, by Formula (\ref{ScFieldDifMap}), yields
\begin{equation} \label{ScFieldEq2}
(I_x \cdot \partial_x) \cdot P
+ \{ [ ( I_x \cdot \grave{\partial}_x ) \cdot \dot{\partial}_x ] \wedge \dot{y} \} \cdot \grave{P}
= (-1)^D \partial_q H
= (-1)^D \partial_q H_{DW} ,
\end{equation}
where the accent has the same role as the dot in that it indicates the scope of differentiation of the corresponding differential operator.

Now, ``dotting" Eq. (\ref{ScFieldEq1}) with $\rev{E}_j$, where $E_j \equiv I_x e_j e_y$, so that $\rev{E}_j \cdot I_x = 0$, we find, after little (geometric) algebra
\begin{equation} \label{ScFieldDWEq1}
e_j \cdot \partial_x \, e_y \cdot y = \rev{E}_j \cdot \partial_P H_{DW} ,
\end{equation}
while Eq. (\ref{ScFieldEq2}) multiplied by $e_y$ reads
\begin{equation} \label{ScFieldDWEq2}
e_j \cdot \partial_x E_j \cdot P = - e_y \cdot \partial_q H_{DW} .
\end{equation}
(Summation from $1$ to $D$ over a repeated indices is implied here and below.) Notice that Eqs. (\ref{ScFieldDWEq1}) and (\ref{ScFieldDWEq2}) are identical to the De Donder-Weyl equations of motion, while $H_{DW}$ is identified with the De Donder-Weyl Hamiltonian \cite{DeDonder,Weyl,Kanat1999,Struckmeier}.

Taking specifically 
\begin{equation} \label{ScFieldHamDW}
H_{DW} = \frac{1}{2} \sum_{j=1}^D (P \cdot E_j)^2 + V(\phi) ,
\end{equation}
where $\phi \equiv e_y \cdot y$, we obtain from Eq.~(\ref{ScFieldDWEq1}) the relation
\begin{equation} \label{ScFieldMomenta}
e_j \cdot \partial_x \phi = \sum_{k=1}^D (\rev{E}_j \cdot E_k)(P \cdot E_k) = P\cdot E_j ,
\end{equation}
which substituted in Eq. (\ref{ScFieldDWEq2}) yields
\begin{equation} \label{ScFieldEOM}
\partial_x^2 \phi = - \partial_\phi V(\phi) .
\end{equation}
This is the equation of motion of a scalar field described by the Lagrangian~(\ref{ScFieldLagr}).

In addition, taking inner product of Eq.~(\ref{ScFieldEq2}) and a spacetime vector $e_j$ provides an interesting relation analogous to the energy conservation, Eq.~(\ref{NonRelEnergyCons}), of non-relativistic mechanics. We have
\begin{equation}
[ e_j \wedge (I_x \cdot \partial_x) ] \cdot P 
+ \{ e_j \wedge [ ( I_x \cdot \grave{\partial}_x ) \cdot \dot{\partial}_x ] \wedge \dot{y} \} \cdot \grave{P}
= (-1)^D e_j \cdot \partial_q H_{DW} ,
\end{equation}
and introducing the resolution $P = \rev{I}_x I_x \cdot P + \rev{E}_k E_k \cdot P$ we get
\begin{equation}
[ e_j \wedge (I_x \cdot \partial_x) ] \cdot \rev{I}_x \, I_x \cdot P 
+ \{ e_j \wedge [ ( I_x \cdot \grave{\partial}_x ) \cdot \dot{\partial}_x ] \wedge \dot{y} \} \cdot \rev{E}_k\, E_k \cdot \grave{P}
= (-1)^D e_j \cdot \partial_q H_{DW} .
\end{equation}
After some algebra we obtain (recall that $\rev{E}_j = e_y e_j \rev{I}_x$)
\begin{equation} \label{ScFieldEnergyMom1}
- e_j \cdot \partial_x P \cdot I_x
+ (-1)^D \{ e_j \wedge [ ( I_x \cdot \grave{\partial}_x ) \cdot (\partial_x \phi) ] \} \cdot (e_k \rev{I}_x) E_k \cdot \grave{P}
= e_j \cdot \partial_q H_{DW} .
\end{equation}
The second term on the left-hand side is simplified as follows:
\begin{align}
&(-1)^D \{ e_j \wedge [ ( I_x \cdot \grave{\partial}_x ) \cdot (\partial_x \phi) ] \} \cdot (e_k \rev{I}_x) E_k \cdot \grave{P} = 
\nonumber\\
&= \{ e_j \wedge e_k \wedge [ ( I_x \cdot \grave{\partial}_x ) \cdot (\partial_x \phi) ] \} \cdot \rev{I}_x\, E_k \cdot \grave{P}
\nonumber\\
&= ( e_j \wedge e_k ) \cdot \{ [ I_x \cdot \big(\grave{\partial}_x  \wedge (\partial_x \phi)\big) ] \cdot \rev{I}_x \} E_k \cdot \grave{P}
\nonumber\\
&= ( e_j \wedge e_k ) \cdot \big(\grave{\partial}_x  \wedge (\partial_x \phi)\big)  E_k \cdot \grave{P}
\nonumber\\
&= ( e_j \wedge e_k ) \cdot \big(\grave{\partial}_x  \wedge (\partial_x \grave{\phi})\big)  E_k \cdot \grave{P} 
\nonumber\\
&= - e_j \cdot \partial_x [ (e_k \cdot \partial_x \phi) \, (E_k \cdot P) ]
+ e_k \cdot \partial_x [ (e_j \cdot \partial_x \phi) \, (E_k \cdot P) ] ,
\end{align}
where the second last equality is a consequence of commutativity of partial derivatives.

Using in Eq. (\ref{ScFieldEnergyMom1}) the Hamiltonian constraint (\ref{CanEOM3}), with $H$ given by (\ref{ScFieldHamConstr}), we arrive at
\begin{equation}
e_k \cdot \partial_x [\delta_{j k} H_{DW} - \delta_{j k} (e_i \cdot \partial_x \phi) (E_i \cdot P) + (e_j \cdot \partial_x \phi) (E_k \cdot P) ]
= e_j \cdot \partial_q H_{DW} .
\end{equation}
For $H_{DW}$ defined in (\ref{ScFieldHamDW}) we can eliminate the momenta, owing to relation (\ref{ScFieldMomenta}), and obtain
\begin{equation}
e_k \cdot \partial_x [ - \delta_{j k} \mathcal{L}(\phi,\partial_x \phi) + (e_j \cdot \partial_x \phi) (e_k \cdot \partial_x \phi) ]
= e_j \cdot \partial_q H_{DW} = 0 .
\end{equation}
The expression in square brackets is nothing but the canonical energy-momentum tensor $\mathcal{T}_{j k}$ of the scalar field with Lagrangian (\ref{ScFieldLagr}), so we finally arrive at the continuity equation
\begin{equation} \label{ScFieldContEq}
e_k \cdot \partial_x \mathcal{T}_{j k} = 0 .
\end{equation}

Note that the equation of motion for the scalar field, Eq.~(\ref{ScFieldEOM}), and the continuity equation (\ref{ScFieldContEq}) for its energy-momentum tensor have common origin in the canonical equations of motion (\ref{CanEOM}). Intuitively, the equation of motion is a result of variations of the surface $\gamma$ in the $y$-direction, while the continuity equation follows from variations in the $x$-plane (see Fig.~\ref{fig:ScField}).

Finally, let us say a few words about the Hamilton-Jacobi theory. Equation~(\ref{HJeq}) with Hamiltonian~(\ref{ScFieldHamConstr}) reads
\begin{equation} \label{ScFieldHJ}
I_x \cdot (\partial_q \wedge S) + H_{DW}(q,\partial_q \wedge S) = 0 ,
\end{equation}
where $S(q)$ is a multivector of grade $D-1$. For concreteness, consider $H_0$ given by Eq.~(\ref{ScFieldHamDW}), and assume $S$ is parallel to $I_x$. Defining the vector $s(q) \equiv S(q) \cdot I_x$, also parallel to $I_x$, we observe that
\begin{align}
I_x \cdot (\partial_q \wedge S) &= \partial_q \cdot s ,
\nonumber\\
E_j \cdot (\partial_q \wedge S) &= e_y \cdot \partial_q \, e_j \cdot s .
\end{align}
Eq.~(\ref{ScFieldHJ}) then takes the form (note that $\partial_\phi s = e_y\cdot\partial_q s$)
\begin{equation}
\partial_q \cdot s + \frac{1}{2} (\partial_\phi s)^2 + V(\phi) = 0 ,
\end{equation}
which coincides with the field-theoretic Hamilton-Jacobi equation derived formerly by Weyl \cite{Weyl}.

\subsection{String theory} \label{sec:ExString}

Probably the simplest nontrivial Hamiltonian to consider is 
\begin{equation}
H = \frac{1}{2}(|P|^2 - \Lambda^2) ,
\end{equation}
where $\Lambda >0$ is a scalar constant, and $|P|$ is the magnitude of $P$ (see definition~(\ref{GAmagnitude})).

According to Formula (\ref{GCderivP2}), the first canonical equation (\ref{CanEOM1}) takes the form 
\begin{equation} \label{StringEOM1}
d\Gamma = \lambda \rev{P} ,
\end{equation}
which substituted into the Hamiltonian constraint (\ref{CanEOM3}) fixes the absolute value of the Lagrange multiplier $\lambda$,
\begin{equation} \label{StringEOM3}
|d\Gamma| = |\lambda| \Lambda .
\end{equation}
Furthermore, substituting Eq.~(\ref{StringEOM1}) into the second canonical equation of motion (\ref{CanEOM2}), dividing by $\lambda$, and using Eq.~(\ref{StringEOM3}) we find
\begin{align} \label{StringEOM}
I_\gamma \cdot \partial_q \, I_\gamma &= 0 ~~~~~ (D=1) ,
\nonumber \\
(I_\gamma \cdot \partial_q ) \cdot I_\gamma &= 0 ~~~~~ (D>1),
\end{align}
where $I_\gamma \equiv d\Gamma / |d\Gamma|$ is the unit pseudoscalar of the surface $\gamma$. This equation has a simple geometric interpretation. It entails vanishing of the mean curvature of $\gamma$, or of its generalization, the \emph{spur} vector (see Ref.~\cite[Ch. 4-4]{Hestenes}).

Eqs.~(\ref{StringEOM1}) and (\ref{StringEOM3}) allow us to rewrite the action (\ref{Action}) in terms of $d\Gamma$,
\begin{equation} \label{NambuGotoAction}
\int_\gamma P \cdot d\Gamma 
= \int_\gamma \frac{1}{\lambda} |d\Gamma|^2
= \pm\,\Lambda \int_\gamma |d\Gamma| ,
\end{equation}
where ``$\pm$" is the sign of $\lambda$, and $|d\Gamma|\equiv\sqrt{d\rev{\Gamma} \cdot d\Gamma}$.
This is the Euclidean Nambu-Goto action of the string theory \cite{Zwiebach}. It is proportional to the volume of the worldsheet $\gamma$, with $\Lambda$ playing the role of string tension (the speed of light is set to unity). Extremals of this action, i.e., solutions of Eq.~(\ref{StringEOM}), minimize the volume, and so are called \emph{minimal surfaces} in mathematical literature \cite{Osserman}. It is worthwhile to mention that the Nambu-Goto string can be formulated also within the De Donder-Weyl Hamiltonian theory (see Ref.~\cite{Kanat1998}, which uses the language of differential forms).


In this example the Hamilton-Jacobi equation (\ref{HJeq}) takes a particularly compact form (cf. Refs.~\cite{Nambu1980} and \cite[Ch.~7]{Kastrup})
\begin{equation} \label{StringHJ}
| \partial_q \wedge S | = \Lambda .
\end{equation}

From now on, let us focus on the case $D=1$, which describes the relativistic particle in the Euclidean spacetime. We will present two methods for finding the physical motions. 

First, suppose that two points, $q_0$ and $q$, lie on $\gamma_{\rm cl}$, multiply Eq.~(\ref{StringEOM}) by $|d\Gamma|$, and integrate along $\gamma_{\rm cl}$ from $q_0$ to $q$. The Fundamental theorem of calculus (\ref{GCintFundThm}) implies that
\begin{equation}
I_\gamma(q) - I_\gamma(q_0) = 0 ,
\end{equation}
i.e., $I_\gamma$ is constant along a physical motion, and $\gamma_{\rm cl}$ are therefore straight lines in $\mathcal{C}$, 
\begin{equation}
\gamma_{\rm cl} = \{ q = v \tau + q_0 \,|\, \tau \in \mathbb{R} \}
\end{equation}
where $q_0 \in \mathcal{C}$ and $v$ is arbitrary constant vector.

Second method utilizes a family of solutions of the Hamilton-Jacobi equation (\ref{StringHJ}), for example,
\begin{equation}
S(q;q_0) = \Lambda |q-q_0| .
\end{equation}
According to Formula (\ref{HJconserved}), derivative of $S$ with respect to the parameters $q_0$ yields  conserved quantities
\begin{equation}
\partial_{q_0} S = - \Lambda \frac{q-q_0}{|q-q_0|} .
\end{equation}
Physical motion are then obtained readily,
\begin{equation}
\gamma_{\rm cl} = \left\{
q \, \bigg| \, \frac{q-q_0}{|q-q_0|} = v
\right\} ,
\end{equation}
where $v$ is an arbitrary constant unit vector.

\section{Conclusion and outlook} \label{sec:Conclusion}

In this article we elaborated on the formulation of classical field theory presented in \cite[Ch.~3]{RovelliQG}, which is based on the notion of partial observables, and on the Hamiltonian constraint. The latter is a function of configuration-space point and the generalized multivector-valued momentum. Starting from the variational principle of Section \ref{sec:VarPrinc} we derived canonical equations of motion (\ref{CanEOM}). We also deduced local Hamilton-Jacobi equation (\ref{HJeq}), which can be a useful tool to find the physical motions.

With three ensuing examples we showed how non-relativistic mechanics, scalar field theory, and string theory can be described in one unifying framework by appropriately selecting the Hamiltonian constraint. In particular, we noticed that equations of motion, and the continuity equation for the energy-momentum tensor (which reduces to the energy conservation equation in the case of non-relativistic mechanics) are in fact of the same origin in the Hamiltonian constraint formalism. Therefore, this formalism may be of interest even for theories that do not assume symmetry between time and space, or spacetime and fields. 

Although we restricted our attention to Euclidean space, an extension of the formalism to pseudo-Euclidean spaces should be relatively straightforward \cite[Ch.~1-5]{Hestenes}. Then, one has to mind, and keep track of, possible sign differences between the reversion and the inversion of unit multivectors. For example, in general, $I_x^{-1} \neq \rev{I}_x$.


Hamiltonian formalism is especially important when it comes to quantization. In particle mechanics, momentum is promoted to a differential operator, and the Schr\"{o}dinger equation is postulated. What is the quantum operator corresponding to the multivector-valued generalized momentum of the Hamiltonian constraint approach? And what does the Schr\"{o}dinger equation look like, once we know the classical Hamilton-Jacobi equation (\ref{HJeq}). Although these questions have not been addressed in general, let us note that there have been studies of quantization in the De Donder-Weyl Hamiltonian theory, where the quantization of momenta is based on generalized Poisson brackets, and a field-theoretic generalization of the Schr\"{o}dinger equation is proposed that features a Clifford-valued wave function, and reduces to the De Donder-Weyl Hamilton-Jacobi equation in the classical limit \cite{Kanat1999,Kanat2013}.

\subsection*{Acknowledgement}
The author would like to thank Igor Kanatchikov for valuable discussions, and the following institutions for financial support: Grant Agency of the Czech Technical University in Prague, Grant SGS13/217/OHK4/3T/14, Czech Science Foundation (GA\v{C}R), Grant GA14-07983S, and Deutsche Forschungsgemeinschaft (DFG), Grant KL 256/54-1.

\appendix
\section{Geometric algebra and calculus} \label{sec:GAGC}

We give a brief introduction into the formalism of geometric algebra and calculus in a way that respect the requirements of this article. For a thorough  and rigorous treatment, the reader is advised to consult monograph \cite{Hestenes}, which we shall frequently quote. Complementary to this is the textbook \cite{DoranLas}, which provides, apart from a pedagogical introduction into the mathematical formalism, many diverse physical applications.

\subsection{Geometric algebra}

Let us start with an $n$-dimensional real vector space $V$, and define the \emph{geometric product} of vectors by the following axioms:
\begin{align} \label{GAaxioms}
\forall a, b, c \in V: 
& ~1)~ a(bc) = (ab)c = abc \nonumber\\
& ~2)~ a(b+c) = ab+ac \nonumber\\
& ~3)~ a^2 > 0 ~~~{\rm for~nonzero}~ a .
\end{align}
This product induces an associative algebra over vector space $V$ --- the \emph{geometric (or Clifford) algebra} $\mathcal{G}(V)$. Frequently, the term ``Clifford algebra" can be encountered in literature. Nevertheless, we prefer the name ``geometric algebra" used originally by Clifford, and advocated by Hestenes \cite{Hestenes} to emphasize its geometric interpretation. Let us note that although it is possible to represent vectors in the algebra by Dirac gamma matrices, it is in fact not very useful, since it provides no insight into the properties of the algebra, neither it simplifies any calculations.

The last axiom in (\ref{GAaxioms}) has far-reaching consequences. Expanding $(a+b)^2 = a^2 + b^2 + a b + b a$, we observe that
\begin{equation} \label{GAinnerprod}
a \cdot b := \frac{1}{2}(a b + b a)
\end{equation}
is a \emph{scalar}. The remaining part of the geometric product,
\begin{equation} \label{GAouterprod}
a \wedge b := \frac{1}{2}(a b - b a) ,
\end{equation}
is a \emph{bivector}. The geometric product of two vectors is therefore decomposed into two parts: symmetric non-associative inner product (\ref{GAinnerprod}), and antisymmetric associative outer product (\ref{GAouterprod}).

The scalar $a \cdot b$ is identified with the scalar product of vectors $a$ and $b$. Positive definiteness of this scalar product follows from the strict inequality in the last axiom in (\ref{GAaxioms}). One could also consider indefinite quadratic forms, but we do not deal with them in this text.

The bivector $a \wedge b$ represents an oriented parallelogram spanned by the two vectors. In fact, it represents a whole equivalence class of parallelograms, since, e.g., $a \wedge b = a \wedge (b + \lambda a)$ for arbitrary scalar $\lambda$, as follows from the antisymmetry of the ``$\wedge$"-product.

Successive multiplication of vectors generates the entire geometric algebra. General elements are called \emph{multivectors}. They decompose into a sum of terms with different \emph{grade}. An element of the algebra has grade $r$ (it is an $r$-\emph{vector}), if it can be written as an exterior product of $r$ vectors, in which case it is called $r$-\emph{blade}, or as a linear combination of such terms. Because $A_r\equiv\blade{a}{r}$ vanishes if and only if the vectors $a_j$ are linearly dependent, it represents an object with dimensionality $r$, namely, the parallelotope spanned by the vectors $a_j$, whose orientation is specified by the order in which the vectors appear in the exterior product. By the Gram-Schmidt process \cite[Ch. 1-3]{Hestenes} it can be shown that every blade $A_r = a_1 \wedge \ldots \wedge a_r$ can be written as a scalar multiple of a geometric product of orthonormal vectors: $A_r = \alpha\, e_1 \ldots e_r$, where $e_j \cdot e_k = \delta_{j k}$. 

General multivector is a linear combination of terms with increasing grade:
\begin{equation}
1 ~~~,~~~ e_j ~~~,~~~ e_j e_k ~(j<k) ~~~,~~~ \ldots ~~~,~~~ e_1 \ldots e_n ,
\end{equation}
where $j,k,\ldots = 1,\ldots,n$. Geometric algebra $\mathcal{G}(V)$ is therefore the linear span of these terms. A useful convention states that scalars are denoted by Greek letters $\alpha, \beta, \ldots$, vectors are denoted by lower case Latin letters $a,b,\ldots$, and other multivectors are denoted by capital letters $A,B,\ldots$. Note, however, that in physical applications there can appear exceptions to these rules due to conventions used in physics. For example, the Hamiltonian in (\ref{HamConstraint}) is denoted by $H$ although it is a scalar function.

When a multivector has some definite grade $r$, this fact is indicated by a subscript, e.g., we write $A_r$. The geometric product between two such multivectors decomposes into a sum of terms with specific grade,
\begin{equation} \label{GAprodDecomp}
A_r B_s = \gr{A_r B_s}{|r-s|} + \gr{A_r B_s}{|r-s|+2} + \ldots \gr{A_r B_s}{r+s} ,
\end{equation}
where the symbols $\gr{M}{r}$ denotes the projection onto the grade-$r$ component of $M$. Projection onto the scalar part is abbreviated $\scal{M} \equiv \gr{M}{0}$. Lowest- and highest-grade terms in the series (\ref{GAprodDecomp}) are designated by
\begin{align} \label{GAinnerouter}
A_r \cdot B_s &\equiv 
\begin{cases}
\gr{A_r B_s}{|r-s|} & ~{\rm if}~ r,s>0 \\
0 & ~{\rm if}~ r=0 ~{\rm or}~ s=0 ,
\end{cases} \nonumber\\
A_r \wedge B_s &\equiv \gr{A_r B_s}{r+s} ,
\end{align}
which are again called interior and exterior product, respectively. The inner product is defined to be zero if either of the multivectors is a scalar. If one of the members is a vector, then relations (\ref{GAinnerouter}) reduce to
\begin{align}
a \cdot A_r &= \frac{1}{2} ( a A_r - (-1)^r A_r a ) , \nonumber\\
a \wedge A_r &= \frac{1}{2} ( a A_r + (-1)^r A_r a ) .
\end{align}
Moreover, expressions such as $a \cdot A$ can be expanded in primitive terms using the identity (see \cite[Ch.~1-1]{Hestenes})
\begin{equation} \label{GAexpansion}
a \cdot (\blade{a}{r}) = \sum_{j=1}^r (-1)^{j-1} a \cdot a_j \, a_1 \wedge \ldots \wedge \check{a}_j \wedge \ldots \wedge a_r ,
\end{equation}
where the check marks vectors that are dropped out from the expression.

To avoid overload of brackets, we have adopted the standard convention that inner and outer products have priority before the geometric product. For example,
\begin{equation}
a \cdot A \, b \wedge c = (a \cdot A) (b \wedge c) .
\end{equation}

To every blade $A_r=\blade{a}{r}$ corresponds a subspace of $V$ spanned by the vectors $a_j$. A generic vector $a \in V$ belongs to this subspace if and only if it is a linear combination of $a_j$'s, i.e., if and only if $a \wedge A_r=0$, as follows from antisymmetry of the outer product. Alternatively, the condition $a \wedge A_r=0$ can be restated as $a \cdot A_r = a A_r$. Vectors in the orthogonal complement are characterized by the requirement $a \cdot A_r = 0$ inferred from expansion (\ref{GAexpansion}).

We quote several useful identities derived in \cite[Ch. 1-1]{Hestenes} that enable efficient manipulations:
\begin{subequations} \label{GAident}
\begin{align} \label{GAidentity1}
A_r \cdot B_s &= (-1)^{r(s-1)} B_s \cdot A_r ~~~~~ {\rm for}~ r \leq s,  
\\  \label{GAidentity2}
A_r \wedge B_s &= (-1)^{rs} B_s \wedge A_r 
\\  \label{GAidentity3}
A_r \cdot (B_s \cdot C_t) &= (A_r \wedge B_s) \cdot C_t ~~~~~ {\rm for}~ r+s \leq t ~{\rm and}~ r,s > 0, 
\\ \label{GAidentity4}
(C_t \cdot B_s) \cdot A_r &= C_t \cdot (B_s \wedge A_r) ~~~~~ {\rm for}~ r+s \leq t ~{\rm and}~ r,s > 0, 
\\  \label{GAidentity5}
A_r \cdot (B_s \cdot C_t) &= (A_r \cdot B_s) \cdot C_t ~~~~~ {\rm for}~ r+t \leq s ,
\\  \label{GAidentity6}
a \cdot (A_r \wedge B_s) &= (a \cdot A_r) \wedge B_s + (-1)^r A_r \wedge (a \cdot B_s) ,
\\  \label{GAidentity7}
a \wedge (A_r \cdot B_s) &= (a \cdot A_r) \cdot B_s + (-1)^r A_r \cdot (a \wedge B_s) ~~~~~ {\rm for}~ s \geq r > 1 .
\end{align}
\end{subequations}

To every multivector $A$ is associated a scalar \emph{magnitude} $|A|$ by
\begin{equation} \label{GAmagnitude}
|A|^2 = \scal{\rev{A} A} ,
\end{equation}
where "$~\rev{.}~$" is the operation of \emph{reversion}, defined trivially on vectors, $\rev{a}=a$, and extended to $\mathcal{G}(V)$ by linearity and the requirement
\begin{equation}
\rev{A B} = \rev{B}\rev{A} .
\end{equation}
For a blade $A_r = a_1 \wedge \ldots \wedge a_r$, $|A|$ is indeed the volume of the parallelotope spanned by the vectors $a_j$. Also note that 
\begin{equation}
\rev{A}_r = (-1)^{r(r-1)/2} A_r ,
\end{equation}
and since $\rev{A}_r A_r = \scal{\rev{A}_r A_r}$, the blade $A_r$ has an inverse,
\begin{equation}
A_r^{-1} = \frac{\rev{A}_r}{|A_r|^2} . 
\end{equation}
For unit blades, the inverse is equal to the reverse: $A_r^{-1}=\rev{A}_r$.

Every grade-$r$ multivector gives rise to a scalar-valued function $\alpha$ of $r$ vector variables,
\begin{equation} \label{GAaltForm}
\alpha(b_1,\ldots,b_r) 
= \rev{A}_r \cdot (b_1 \wedge \ldots \wedge b_r) ,
\end{equation} 
which is linear in each argument, and changes sign whenever two vectors are exchanged, i.e., $\alpha$ is an \emph{alternating form}. In fact, every alternating form can be represented by some multivector $A_r$ in this way, and operations on forms can be naturally expressed in terms of operations on the corresponding multivectors \cite[Ch. 1-4]{Hestenes}. 

If $A_r$ is a blade, then the inner product of two $r$-blades, $\rev{A}_r = \revblade{a}{r}$ and $B_r = \blade{b}{r}$, can be expressed as the determinant of the matrix of scalar products $a_j \cdot b_k$,
\begin{equation} \label{GAdet}
\rev{A}_r \cdot B_r 
= \det (a_j \cdot b_k) .
\end{equation}
In addition, using the expansion (\ref{GAexpansion}) we derive
\begin{equation} \label{GAminor}
(\rev{A}_r \cdot a) \cdot (b \cdot B_r)
= \sum_{l,m=1}^n (-1)^{l+m} (a \cdot a_l)  (b \cdot b_m) \,
{\rm minor}(a_j \cdot b_k | l,m) ,
\end{equation}
where
\begin{equation}
{\rm minor}(a_j \cdot b_k | l,m) 
\equiv (a_r \wedge \ldots \wedge \check{a}_l \wedge \ldots \wedge a_1) \cdot (b_1 \wedge \ldots \wedge \check{b}_m \wedge \ldots \wedge b_r)
\end{equation}
denotes the $(l,m)$ minor of the matrix $a_j \cdot b_k$, i.e., the determinant of the latter matrix with the $l$th row and $m$th column erased.

Let us choose an orthonormal basis $\{ e_j \}_{j=1}^{n}$ of the vector space $V$. Any vector $a$ can be expanded as
\begin{equation} \label{GAVectExpand}
a = \sum_{j=1}^n (a \cdot e_j) e_j ,
\end{equation}
where $a \cdot e_j$ are the components of $a$ with respect this basis. Moreover, any grade-$r$ multivector $A_r$ can be expressed as a sum
\begin{equation} \label{GAMultivExpand}
A_r = \sum_{|J|=r} (A_r \cdot \rev{e}_J) e_J = \sum_{|J|=r} (A_r \cdot e_J) \rev{e}_J 
\end{equation}
over all ordered sets of indices $J=(j_1,\ldots,j_r)$, $j_1<\ldots<j_r$, where $e_J \equiv e_{j_1}\ldots e_{j_r}$. Relation (\ref{GAMultivExpand}) can be proved with a help of the formula (see Eq.~(3.14) in \cite[Ch.~1-3]{Hestenes})
\begin{equation} \label{GAMultiDelta}
\rev{e}_K \cdot e_J 
= (e_{k_r} \ldots e_{k_1}) \cdot (e_{j_1} \ldots e_{j_r})
= \delta_{j_1}^{k_1} \ldots \delta_{j_r}^{k_r}
\equiv \delta_J^K .
\end{equation}

\subsection{Geometric calculus: Differentiation}

We will now move towards the \emph{geometric calculus}, that is, the theory of differentiation and integration developed by D. Hestenes \cite{Hestenes}, which takes advantage of the rich algebraic structure of geometric algebra.

Our setting involves a Euclidean vector space $V$, corresponding to the configuration space $\mathcal{C}$ of partial observables, which holds at every point $q \in V$ a copy of geometric algebra $\mathcal{G}(V)$. Take function $F(q)$ with values in $\mathcal{G}(V)$, and a vector $a$. The \emph{derivative} of $F$ in the \emph{direction} $a$ is defined in the usual manner,
\begin{equation}
a \cdot \partial_q F(q) := \lim_{\eps \rightarrow 0} \frac{F(q+\eps a) - F(q)}{\eps} .
\end{equation}
The \emph{vector derivative} of $F$ is defined with the help of an orthonormal basis $\{ e_j \}_{j=1}^n$ of $V$,
\begin{equation} \label{GAvectDer}
\partial_q F(q) := e_j (e_j \cdot \partial_q) F(q) 
\end{equation}
(summation over $j$ is implied).
The operator $\partial_q = e_j (e_j \cdot \partial_q)$ has algebraic properties of a vector, and hence we may separate the vector derivative into two parts,
\begin{equation}
\partial_q F = \partial_q \cdot F + \partial_q \wedge F ,
\end{equation}
called \emph{divergence} and \emph{curl}, respectively. When $F$ is vector-valued, we recover the familiar differential operators of vector calculus. For scalar $F$, $\partial_q F$ is simply the \emph{gradient}.

The vector derivative is obviously linear. To express the product rule we employ the ``overdot" notation specifying which function in the product is being differentiated,
\begin{equation}
\partial_q (F G) = \dot{\partial}_q \dot{F} G + \dot{\partial}_q F \dot{G} . 
\end{equation}
The reason is that the vector $\partial_q$ need not commute with the other multivectors. Of course, overdots can always be eliminated by introducing a basis, for example, $\dot{\partial}_q F \dot{G} = e_j F (e_j \cdot \partial_q) G$. Some explicit formulas for vector derivatives of elementary functions can be found in \cite[Ch.~2-1]{Hestenes}. 

Differential forms are skew-symmetric linear functions of differential arguments $dq_1,\ldots,dq_r$ that may vary from point to point. Just like in Eq.~(\ref{GAaltForm}), they can be expressed as
\begin{equation} \label{GAdifForm}
\alpha(\blade{dq}{r};q) = \rev{A}_r(q) \cdot (\blade{dq}{r}) ,
\end{equation}
where $A_r$ is an $r$-vector function (see \cite[Ch.~6-4]{Hestenes}). Exterior derivative of $\alpha$ is tantamount to taking the curl of $A_r$,
\begin{equation} \label{GCformExtDeriv}
d\alpha(\blade{dq}{r+1};q) = \big(\dot{\rev{A}}_r \wedge \dot{\partial}_q \big) \cdot (\blade{dq}{r+1}) .
\end{equation}

Let us now consider a diffeomorphism $f$ that maps points $q \in V$ to $V$, and form the directional derivative
\begin{equation} \label{GAdifMap}
\underline{f}(a;q) \equiv a \cdot \partial_q f(q).
\end{equation}
This gives rise to a $q$-dependent linear function, the \emph{differential} of $f$, mapping vector $a$ to a new vector $\underline{f}(a)$. (In standard differential geometry $\underline{f}$ is called the \emph{push-forward} derived from the diffeomorphism $f$.) It is natural to extend the domain of $\underline{f}$ to general multivectors, demanding linearity and the \emph{outermorphism} property \cite[Ch. 3-1]{Hestenes}
\begin{equation}
\underline{f}(A \wedge B) = \underline{f}(A) \wedge \underline{f}(B) .
\end{equation}

Let $\overline{f}$ denote the \emph{adjoint} of the linear transform $\underline{f}$. It fulfils, for any two vectors $a$ and $b$,
\begin{equation} \label{GCadjoint}
b \cdot \underline{f}(a) = \overline{f}(b) \cdot a 
~~~\Rightarrow~~~
\overline{f}(b;q) = \partial_q f(q) \cdot b ,
\end{equation}
and corresponds to the \emph{pull-back} of differential geometry.
The adjoint is extended to an outermorphism in the same way as the differential,
\begin{equation}
\overline{f}(A \wedge B) = \overline{f}(A) \wedge \overline{f}(B) .
\end{equation}
For scalar arguments we define
\begin{equation}
\underline{f}(\alpha) = \overline{f}(\alpha) = \alpha .
\end{equation}
 
Although the inner product is not, in general, preserved by the differential and adjoint outermorphisms, the following useful relations hold \cite[Ch. 3-1]{Hestenes}
\begin{align} \label{GAadjointIdent}
A_r \cdot \overline{f}(B_s) = \overline{f}[ \underline{f}(A_r) \cdot B_s ] ~~~~~ {\rm for}~ r \leq s , \nonumber\\
\underline{f}(A_r) \cdot B_s = \underline{f}[ A_r \cdot \overline{f}(B_s) ] ~~~~~ {\rm for}~ r \geq s .
\end{align}

Let us investigate the transformations of multivectors under an infinitesimal diffeomorphism $f(q) = q + \delta q(q)$. For a vector we have
\begin{equation}
\underline{f}(a) = a + a \cdot \partial_q \, \delta q .
\end{equation} 
For a blade $A_r = \alpha\, e_1 \wedge \ldots \wedge e_r$, where $\{ e_j \}_{j=1}^n$ is an orthonormal basis of $V$,  we derive
\begin{align} \label{GAinfsmOutDeriv}
\underline{f}(A_r) &= 
\alpha (e_1 + e_1 \cdot \partial_q \, \delta q) \wedge \ldots \wedge (e_r + e_r \cdot \partial_q \, \delta q) \nonumber\\
&\approx A_r + \alpha \sum_{j=1}^r e_1 \wedge \ldots \wedge (e_j \cdot \partial_q \, \delta q) \wedge \ldots \wedge e_r  \nonumber\\
&= A_r + \alpha \sum_{j=1}^r [ (e_1 \wedge \ldots \wedge e_r) \cdot e_j ] \wedge (e_j \cdot \partial_q \, \delta q) ,
\end{align}
where we have used backwards the expansion formula (\ref{GAexpansion}). Now, the sum over $j$ can be extended to run from $1$ up to $n$, since $e_j$'s are orthogonal to $A_r$ for $j > r$. Hence, applying the definition of vector derivative (\ref{GAvectDer}) we arrive at
\begin{equation} \label{GAinfsmOut}
\underline{f}(A_r) \approx A_r + (A_r \cdot \partial_q) \wedge \delta q .
\end{equation}

The Hamiltonian $H(q,P)$ is a function of vector variable $q$ and $D$-vector variable $P$. In order to differentiate the composite function $H(q,P(q))$ with respect to $q$, we need a notion of differentiation with respect to the multivector variable $P$. 

Therefore, suppose $F(P)$ is a multivector-valued function of a grade-$D$ multivector argument $P$, and $A$ is an arbitrary $D$-vector. We define the \emph{$A$-derivative}
\begin{equation}
A \cdot \partial_P F(P) := \lim_{\eps \rightarrow 0} \frac{F(P + \eps A) - F(P)}{\eps} ,
\end{equation}
and the \emph{multivector derivative}
\begin{equation} \label{GCmultivecDeriv}
\partial_P F(P) := \sum_{|J|=D} \rev{e}_J (e_J \cdot \partial_P) F(P) ,
\end{equation}
where $\{ e_J \}_{|J|=D}$ is an orthonormal basis of the subspace of $D$-vectors of the geometric algebra $\mathcal{G}(V)$.

Multivector derivatives of some elementary functions are listed in \cite[Ch. 2-2]{Hestenes}.
For example,
\begin{equation}
A \cdot \partial_P |P|^2 = A \cdot \rev{P} + P \cdot \rev{A} = 2 A \cdot \rev{P} ,
\end{equation}
which implies
\begin{equation} \label{GCderivP2}
\partial_P |P|^2 = 2 \rev{P} .
\end{equation}
The chain rule for differentiation gives
\begin{equation} \label{GCchainH}
a \cdot \partial_q H(q,P(q))
= (a \cdot \dot{\partial}_q) H(\dot{q},P(q)) 
+ (a \cdot \dot{\partial}_q) (\dot{P}(q) \cdot {\partial}_P) H(q,P)|_{P=P(q)} ,
\end{equation}
where the meaning of overdots should be evident.

\subsection{Geometric calculus: Integration}

Consider a $D$-dimensional submanifold $\gamma$ of $V$, whose tangent space at every point $q \in \gamma$ is represented by a unit $D$-blade $I_\gamma(q)$. $I_\gamma$ is called the \emph{unit pseudoscalar of $\gamma$} \cite[Ch. 4-1]{Hestenes}, and it defines also the orientation of $\gamma$. Recall that a vector $a(q)$ is tangent to $\gamma$ if and only if $a(q) \wedge I_\gamma(q) = 0$.

\emph{Directed integral} of a multivector-valued function $F(q)$ over the manifold $\gamma$ is defined in an intuitive way, by approximating $\gamma$ with a chain of simplices $\Delta\Gamma(q)$ (see \cite{Sobczyk}, \cite[Ch.~7]{Hestenes}, or \cite[Ch.~6.4]{DoranLas}), and taking the Riemann sum,
\begin{equation} \label{GCint}
\int_\gamma d\Gamma(q) F(q) := \lim_{n \rightarrow \infty} \sum_{i=1}^n \Delta\Gamma(q_i) F(q_i) .
\end{equation}
Here $d\Gamma = |d\Gamma| I_\gamma$ is the oriented surface element of $\gamma$. \emph{Undirected integral} of $F$ is equal to the directed integral of $I_\gamma^{-1} F$. 

In a more general setting, we can define the directed integral
\begin{equation} \label{GCintGen}
\int_\gamma L(d\Gamma(q);q) := \lim_{n \rightarrow \infty} \sum_{i=1}^n L(\Delta\Gamma(q_i);q_i) ,
\end{equation}
where $L(A_D;q)$ is a multivector-valued function, linear in the $D$-vector argument $A_D$. Obviously, the integral (\ref{GCintGen}) reduces to (\ref{GCint}) if we choose $L(A_D;q) = A_D F(q)$.

Suppose we are given two surfaces $\gamma$ and $\gamma'$, where $\gamma' = \{ q'=f(q) \,|\, q \in \gamma \}$ for a certain diffeomorphism $f$. The unit pseudoscalar $I_\gamma$ is mapped by the outermorphism $\underline{f}$ onto $\gamma'$, where it is proportional to $I_{\gamma'}$. We shall assume that the proportionality constant is positive, i.e., that the surfaces $\gamma$ and $\gamma'$ have common orientation. 

Oriented surface elements are related by $d\Gamma' = \underline{f}(d\Gamma)$. To compare integrals along the two surfaces, we can change the integration variables (see also \cite[Ch. 7-5]{Hestenes}),
\begin{equation} \label{GCintChangeVar}
\int_{\gamma'} L(d\Gamma'(q');q') = \int_\gamma L(\underline{f}(d\Gamma(q);q);f(q)) .
\end{equation}


Finally, let us quote the \emph{Fundamental theorem of geometric calculus} \cite[Ch. 7-3]{Hestenes}, \cite[Ch. 6.5]{DoranLas}, or the generalized Stokes theorem, which relates integral over a surface $\gamma$ and integral over its boundary $\partial \gamma$:
\begin{equation} \label{GCintFundThm}
\int_{\partial\gamma} L(d\Sigma;q) = \int_\gamma L(d\Gamma \cdot \dot{\partial};\dot{q}) ,
\end{equation}
where here the first argument of $L$ is a multivector of grade $D-1$, and $d\Sigma$ is the surface element of the boundary $\partial\gamma$.

\section{Component form of the canonical and Hamilton-Jacobi equations} \label{sec:Comp}

With respect to an orthonormal basis $\{e_j\}_{j=1}^{D+N}$ of the configuration space $\mathcal{C}$, the momentum multivector may be expanded as (see Eq.~(\ref{GAMultivExpand}))
\begin{equation}
P = \sum_{|J|=D} P_J \rev{e}_J \quad,\quad P_J \equiv P \cdot e_J .
\end{equation}
The surface element $d\Gamma$ can be thought of as an outer product of $D$ infinitesimal vectors,
\begin{equation}
d\Gamma 
= \blade{dq}{D} ,
\end{equation}
where each $dq_j$ can be decomposed as $dq_k = dq_k^j e_j$ (summation over the repeated index from $1$ to $D+N$ is implied).
Denoting the derivative in direction $e_j$ by $\frac{\partial}{\partial q_j} \equiv e_j \cdot \partial_q$, and the multivector derivative in direction $\rev{e}_J=e_{j_D}\wedge\ldots\wedge e_{j_1}$ by $\frac{\partial}{\partial P_J} \equiv \rev{e}_J \cdot \partial_P$, the canonical equations of motion (\ref{CanEOM}) are cast equivalently as follows:
\begin{subequations} \label{CanEOMComp}
\begin{align} 
\label{CanEOM1Comp}
\lambda \, \frac{\partial H(q_i,P_I)}{\partial P_J} 
&= \rev{e}_J \cdot (\blade{dq}{D})
= \det(dq^{j_k}_l) , 
\\ \label{CanEOM2Comp}
- \lambda \, \frac{\dot{\partial} H(\dot{q_i},P_I)}{\partial q_j} 
&= \begin{cases}
dq_1 \cdot \partial_q \, P \cdot e_j = dq_1^k \frac{\partial P_j}{\partial q_k} & ~~{\rm for}~ D=1 \\
(d\Gamma \cdot \partial_q) \cdot (e_j \cdot P)
= ( \frac{\partial P}{\partial q_k} \cdot e_j ) \cdot (e_k \cdot d\Gamma) &  ~~{\rm for}~ D>1 ,
\end{cases}
\\
\label{CanEOM3Comp} 
H(q_i,P_I) &= 0 .
\end{align}
\end{subequations}
The determinant in the first equation arises from Formula~(\ref{GAdet}).
The subsequent Formula~(\ref{GAminor}) can be employed to cast the second equation, case $D>1$, as
\begin{align}
- \lambda \, \frac{\dot{\partial} H(\dot{q_i},P_I)}{\partial q_j} 
&= \sum_{|J|=D} \frac{\partial P_J}{\partial q_k} 
\big( (e_{j_D}\wedge\ldots\wedge e_{j_1}) \cdot e_j \big) \cdot \big(e_k \cdot (\blade{dq}{D}) \big) \nonumber\\
&= \sum_{|J|=D} \frac{\partial P_J}{\partial q_k} 
\sum_{l,m=1}^D dq_m^k \, {\rm minor}(dq_s^{j_r}|l,m) (-1)^{l+m} \delta_{j,j_l} .
\end{align}

If we parametrize the surface $\gamma$ by $D$ coordinates $\tau_1,\ldots,\tau_D$, the infinitesimal tangent vectors are expressed
\begin{equation}
dq_j^k = e_k \cdot dq_j = e_k \cdot \frac{\partial q}{\partial\tau_j} d\tau_j \equiv \frac{\partial q_k}{\partial\tau_j} d\tau_j \quad (j~ {\rm not~summed~over}),
\end{equation}
where $q=q(\tau_1,\ldots,\tau_D)$ is the embedding mapping from the parameter space to the configuration space. The corresponding Jacobian appears on the right-hand side of Eq.~(\ref{CanEOM1Comp}), which now reads
\begin{equation}
\det(dq^{j_k}_l) 
= \det\left(\frac{\partial q_{j_k}}{\partial\tau_l} \right) d\tau_1\ldots d\tau_D ,
\end{equation} 
while the infinitesimal element $d\tau_1\ldots d\tau_D$ can be divided out to renormalize the Lagrange multiplier $\lambda$.

Let us now focus on the local Hamilton-Jacobi equation (\ref{HJeq}). In components we have
\begin{equation}
H(q_i,(\partial_q \wedge S) \cdot e_I) = 0 ,
\end{equation}
with
\begin{equation}
(\partial_q \wedge S) \cdot e_I
= (-1)^{D-1} \frac{\partial S}{\partial q_k} \cdot \big(e_k \cdot (e_{i_1}\wedge\ldots\wedge e_{i_D})\big)
= \sum_{j=1}^D (-1)^{D+j} \frac{\partial S_{i_1\ldots\check{i_j}\ldots i_D}}{\partial q_{i_j}} ,
\end{equation}
where $S_{i_1\ldots\check{i_j}\ldots i_D} \equiv S \cdot (e_{i_1}\wedge\ldots\wedge \check{e}_{i_j}\wedge\ldots\wedge e_{i_D})$ are the components of a $D-1$-vector $S$. For example, when $D=2$,
\begin{equation}
(\partial_q \wedge S) \cdot e_I
= \frac{\partial S_{i_1}}{\partial q_{i_2}} - \frac{\partial S_{i_2}}{\partial q_{i_1}} .
\end{equation}

\end{document}